# A NOVEL METHOD OF TACTILE ASSESSMENT OF ARTERIES USING COMPUTATIONAL APPROACH

Ali Abouei Mehrizi*[1], Siamak Najarian[1], Roozbeh Khodambashi[2], Salman Dehkhoda[3]

[1]Faculty of Biomedical Engineering, Amirkabir University of Technology, Tehran, Iran.
[2]Islamic Azad University, Majlesi Branch, Majlesi City, Iran.
[3]Artesh University of Medical Sciences, Imam-Reza Hospital, West Fatemi Avenue, Tehran, Iran.

**Corresponding Author:** Ali Abouei Mehrizi, Research Assistant of Biomechanics, Lab of Artificial Tactile Sensing and Robotic Surgery, Faculty of Biomedical Engineering, Amirkabir University of Technology, Hafez Avenue, Tehran, Iran. Telephone Number: (+98-21)-6454-2378. Fax Number: (+98-21)-6646-8186. E-mail address: abouei.ali@gmail.com.

**ABSTRACT**

In this study, palpating and rolling arteries with different physical properties and tactile distinction of them from each other, using finite element method, are presented. Five models have been created: healthy artery containing blood, axisymmetric stenotic artery containing blood, non-axisymmetric stenotic artery containing blood, artery empty of blood, and filled artery. After solving these models, it was observed that the stress graph of the models included a time-dependent stress peak except for the last two models. The value of graphs peak and difference between their maximum and minimum stresses are considered as the criteria of distinction of the models.

**Key words:** Artery, Stenosis, Palpation, Rolling, Modeling.

## 1. INTRODUCTION

Discriminating between different biological tissues during open surgery is mainly achieved by the surgeon's sense of touch; this sense helps to determine how much force to exert by the hands and how to avoid injuring nontarget tissues [1]. In some procedures, tactile sensations are essential to quickly complete a task. For example, hard lumps in soft tissue are detected by probing the tissue with fingers; arteries are localized during dissection by feeling for a time varying pressure; and the structural integrity of a blood vessel wall is assessed by palpating or rolling it between the fingers [2], [3]. However, in many surgical cases, vascular surgeons may make mistakes during diagnosis of physical properties of arteries. The reasons of these errors are the limitations of human tactile sense and the surgeon's need for high versatility in diagnosis of vascular diseases. Also, minimally invasive surgery (MIS) is now being widely used as one of the most preferred choices for various types of operations [4]-[8]. Despite the advantages of MIS and its growing popularity, it suffers from one major drawback; it decreases the sensory perception of the surgeon who may accidentally cut or incur damage to some of the soft biological tissues. One of the main difficulties encountered in this area is the inability in assessment of human vessels like determining the region of the stenosis of an artery during vascular surgery.

Although many studies have been performed on artery detection, one cannot find any research about artery assessment during surgery. This is because nearly all of these studies have used imaging techniques such as angiography and Doppler ultrasound methods and these techniques are preoperational methods [2]. Artificial tactile sensing is a new technique for assessing blood vessels during open surgery and MIS. Many studies have used artificial tactile sensing technique on examination of soft tissue [9]-[16]. Very few numbers of these studies could be found about artery in which they have used this method and only for detecting a simulated artery in a phantom of soft tissue [17]-[19].

In the present study, modeling of different blood vessels was carried out to pattern out surgeon's palpation by using the finite element method (FEM). Therefore, five models have been created: healthy artery containing pulsatile liquid, axisymmetric stenotic artery containing pulsatile liquid, non-axisymmetric stenotic artery containing pulsatile liquid, artery without liquid, and filled artery. Then, these models were solved and the stress graph of the contact area between the finger and the vessels was extracted. Due to the dependency of these graphs to time, the value of their peak, and the variation of these values during time solution, these models can be differentiated from each other. Finally, the rolling of the five vessels by surgeon's fingers was modeled and the importance of this work in the surgeon's diagnosis was described.

## 2. MATERIALS AND METHODS

Upon physical contact between the surgeon's finger and the human tissues, some parameters of touch can be used as a criterion for diagnosis. This criterion can be force, pressure (stress), displacement (strain), temperature, humidity, roughness, stiffness, and softness that appeared on the surface of the touched tissue. In this study, we consider the stress as the criterion of detection during palpation of the tissue by fingers.

**Definition of Problem**

Physicians and surgeons use their fingers to feel target tissue to estimate its size, shape, firmness, or location during touch process [1]. For example, vascular surgeons generally employ palpating and rolling of arteries during surgery for diagnosis of vascular disease [2]. In this regard, we considered a phantom of a bifurcated artery as shown in Fig. 1-*a*. Having imitated palpation and modeled it using a computer, changes appearing on the bottom side of the finger in several locations of the artery were studied.

**Simplifications and Assumptions**

According to the physical standard for soft tissue simulation [20], circular tubes were chosen as the simplified model of the bifurcated artery. We considered five specific sites of the artery to palpate, roll, and assess its conditions. The cross section of these five locations was supposed as five distinct models, as illustrated in Fig. 1-*b*. Here, we introduce and explain the models, i.e., the five cross sections. Cross sections A-A, B-B, C-C, D-D, and E-E indicate a healthy artery containing blood, a 30% axisymmetric stenotic artery containing blood, a 30% non-axisymmetric stenotic artery containing blood, an artery without blood, and a filled artery (100% stenotic artery), respectively.

In this modeling, arteries were assumed to be completely circular. The inner radius of the arteries is 2.25 mm. The mean wall thickness of the arteries was considered 0.315 mm. The tissue and the bone of the finger and the artery were assumed elastic and isotropic with a modulus of elasticity of 100, 50000, and 400 kPa [21], [22], respectively. The modulus of elasticity of the stenosis was assumed to be the as same the artery. Poisson's ratio for all materials was considered 0.45. The blood pressure waveforms of the brachial artery were applied for modeling the inner pressure of the artery [23].

**Finite Element Modeling**

This problem was modeled and solved by the numerical method of finite element analysis, using ABAQUS software (Release 7.6.1). For simulating the palpation process of the models, e.g., cross section A-A, first, we defined a contact between the two simplified fingers and the artery. Then, a constant displacement of -3 mm in direction *y* was applied to the top line of the moveable finger during 1 second of the solution time (see Fig. 2). Afterwards, while holding the moveable finger, the blood pressure was applied to the inner surface of the artery during 0.83 second. To complete the simulation of palpation effect, the bottom line of the fixed finger was fixed in the direction of finger displacement (see Fig. 2-c). The models were meshed with CPS4R (a 4-node bilinear plane stress quadrilateral), which is well suited to model regular meshes. Finally, we solve these models for duration of 1.83 seconds.

In addition, we statically modeled and simulated the rolling of the arteries by a surgeon's simulated finger to obtain beneficial information on their physical and mechanical properties. For the process of approaching, we determined a suppositional vertical line stick to the artery as shown in Fig. 3. Then, for every step of solution, we rotated the arteries 45 degrees to get 360 degrees rotation of the arteries. In Fig. 3, we illustrated the half of the rolling process of cross section C-C. This rolling process was created for four other cross sections.

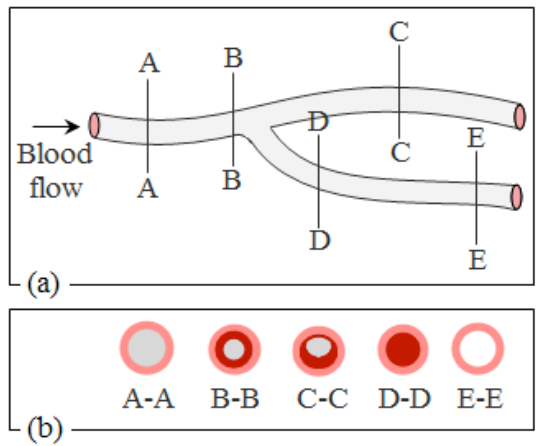

**Fig. 1.** (a) The phantom of a bifurcated artery and (b) The cross section of the five indicated situations.

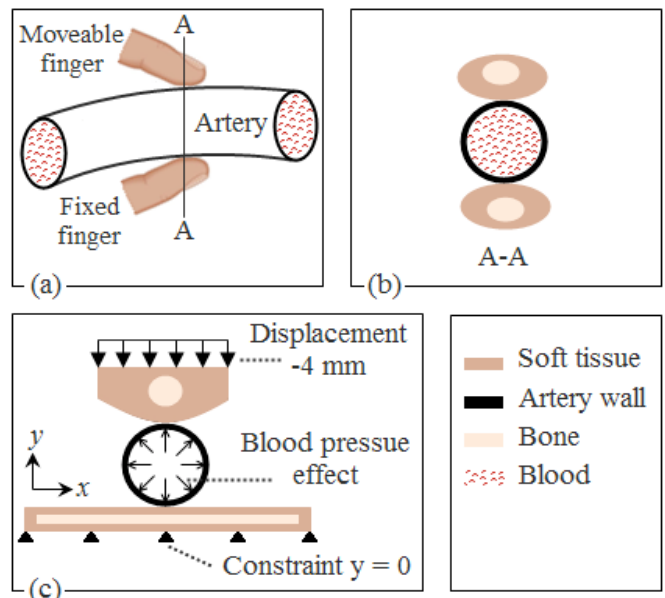

**Fig. 2.** (a) Schematic of palpating location A by a finger, (b) The cross section A-A, and (c) The simplified finite element model of cross section A-A. This finite element model was created for four other cross sections.

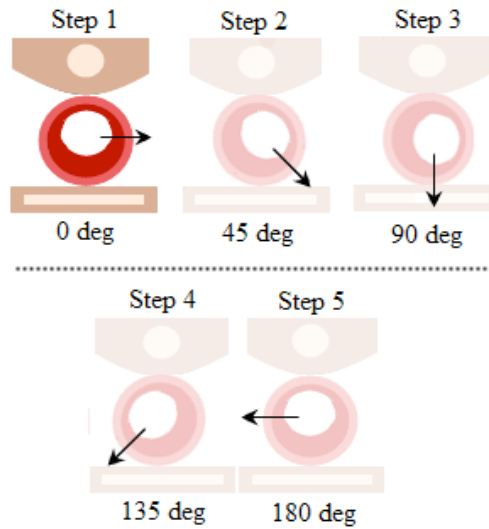

**Fig. 3.** The half of rolling process of a 30% non-axisymmetric stenotic artery containing pulsatile liquid (cross section C-C).

## 3. RESULTS

According to the procedure of this study, we extracted the von Mises stress values of the bottom line of the finger in the finite element models at specific times 1.1 (systolic time, maximum blood pressure), 1.3 (arbitrary time), and 1.4 (diastolic time, minimum blood pressure) seconds as shown in Fig. 4. We chose the bottom line of the finger for its role in transfer of tactile data in real palpation. It was observed that the stress graphs of the five models were not the same. The amount of their stress peak at a specific time 1.1 second, the dependency of stress graphs to time, and the difference of the maximum and minimum stress peaks of each model are as our criteria in the distinction of the models. Meanwhile, the cross section A-A which belongs to the healthy part of the bifurcated artery, was considered as our reference to distinguish these models from each other. We assumed the difference of the maximum and minimum stress as the intensity of arterial pulse or pulse intensity.

Based on Fig. 4, it is obvious that the stress graphs *a*, *b*, and *c* depend on time but the stress graphs *d* and *e* are time independent. The quantity of stress peak at time 1.1 second for the stress graphs *a* to *e* are 23.78, 30.17, 25.13, 45.77, and 6.41 kPa, respectively. Also, the pulse intensity of the stress graphs *a* to *e* are 7.44, 4.31, 2.97, 0, and 0 kPa, respectively.

Also, as shown in Fig. 5, we extracted the maximum values of the contact stress of the finger in nine angular steps of the rolling process for the models at three specific times. As presented in Fig.5, only the stress graph of cross section C-C has a peak in each time. The difference of stress graphs in other cross sections is related to their dependency on time and their maximum stress value.

## 1. DISCUSSION AND CONCLUSION
**Discussion**

In this study, palpating and rolling of five locations of a bifurcated diseased artery were simulated by finite element method. The purpose of current study is distinguishing of the five locations according to three criteria as mentioned before. The extracted results are the contact stresses of the bottom line of the finger at three times in the five models. From Fig. 4, it is evident that the stress graphs *a* to *c* are time dependent and the summit value of them at 1.1 second is 23.78, 30.17, and 25.13, respectively. Also, their pulse intensity is 7.44, 4.31, and 2.96 kPa, respectively. According to the stress graph of the reference cross section that is A-A, it is can be derived that the stress graphs *b* and *c* belong to diseased parts of the artery. This is because their pulse intensity is lower and their summit at 1.1 second is higher than the healthy part of the artery. The summit value of stress graph *d* at 1.1 second as compared to the summit value of other ones is the highest, i.e., 45.77 kPa. In addition, the summit is time independent. Therefore, stress graph *d* belongs to the filled part of the artery. In other words, the artery has 100% stenotic in location D-D. Since the stress graph *e* does not depend on time and its summit value is the lowest among the cases, this graph implies that we have a healthy part of the artery which is empty of blood. This means this graph belongs to cross section E-E.

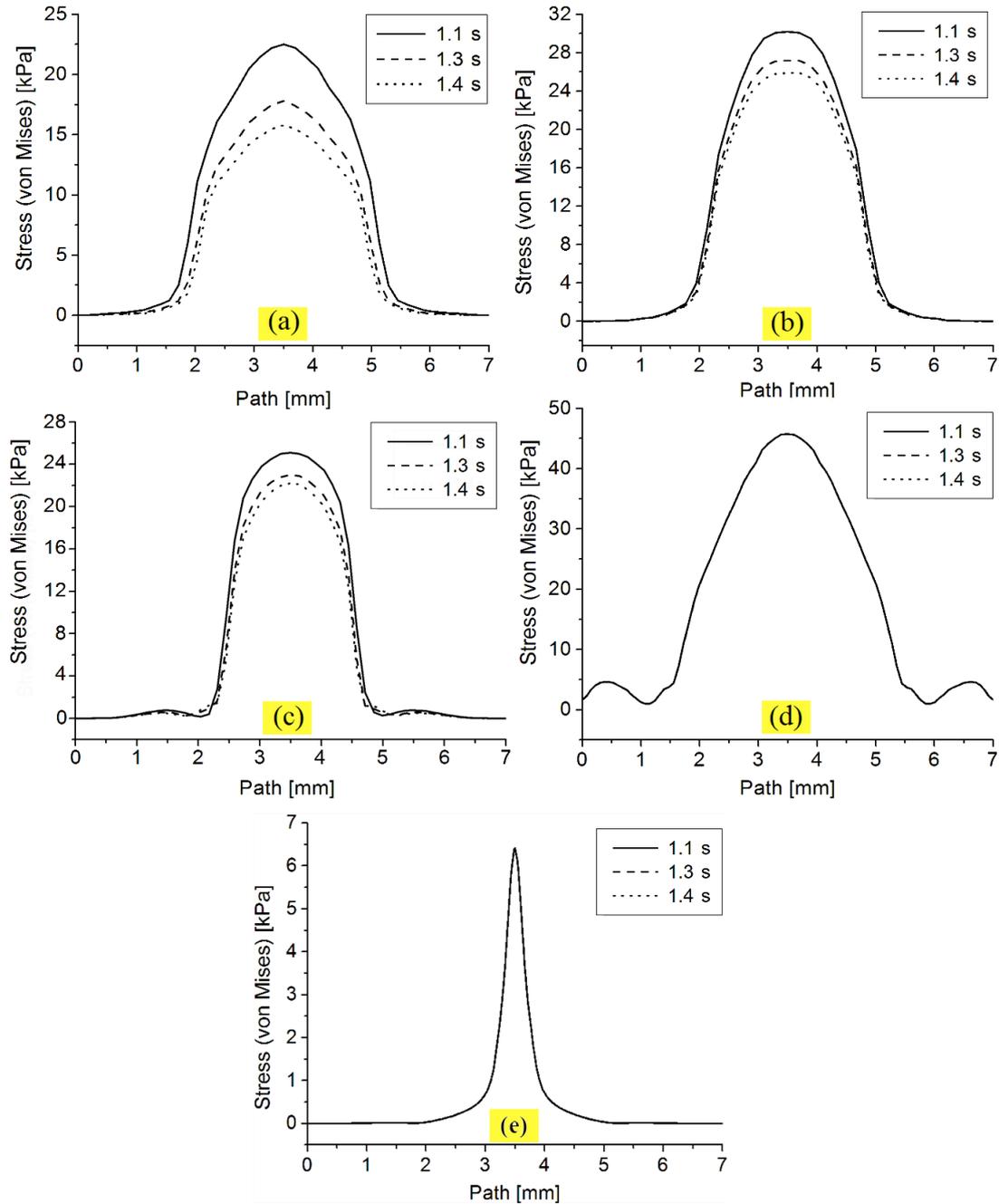

**Fig. 4.** The stress graph of bottom side of finger in models of cross sections (a) A-A, (b) B-B, (c) C-C (at situation 0 degree), (d) D-D, and (e) E-E.

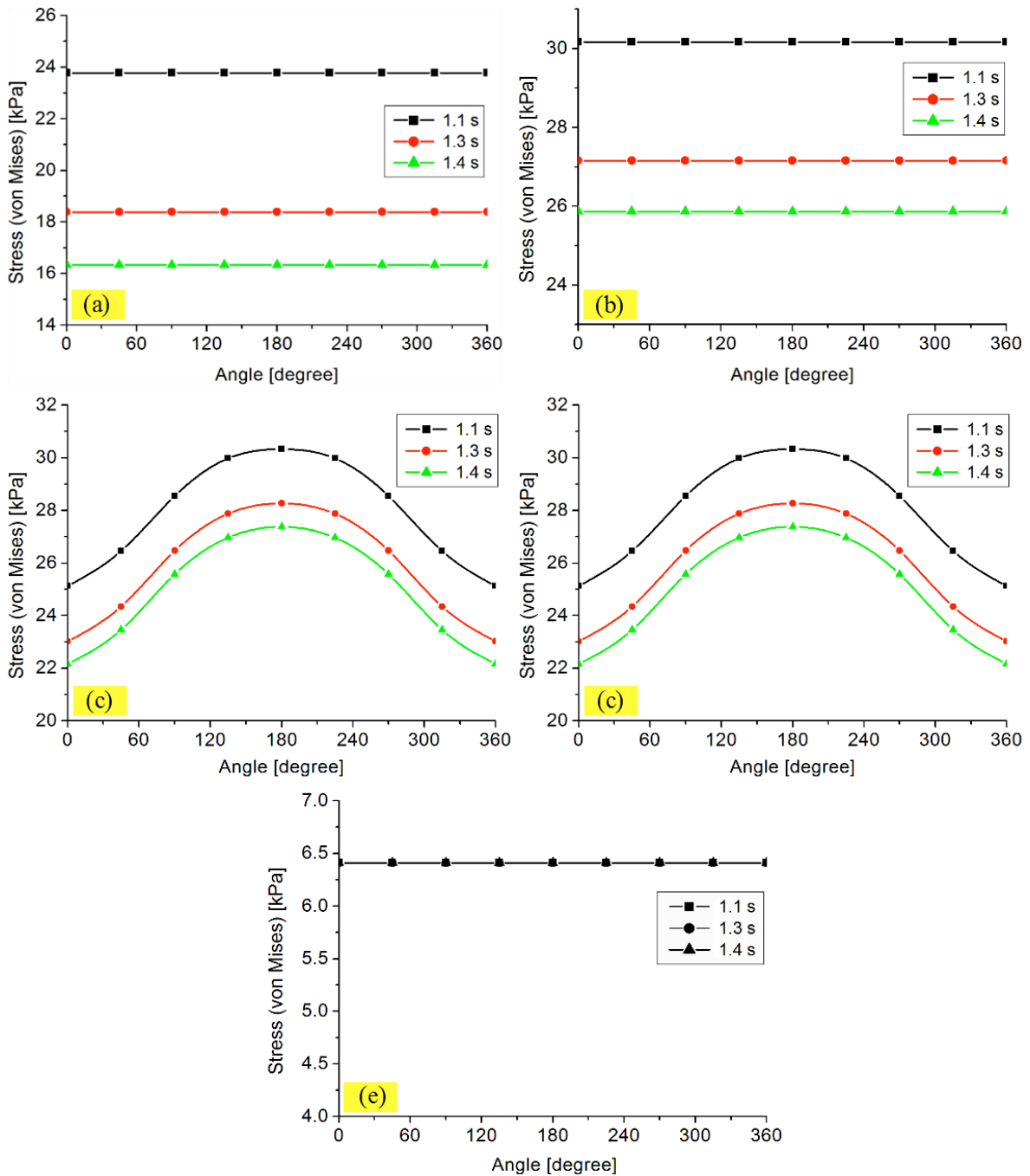

**Fig. 5.** The stress graph related to the rolling of five models of cross sections (a) A-A, (b) B-B, (c) C-C (at situation 0 degree), (d) D-D, and (e) E-E.

We could not differentiated between stress graphs *b* and *c* through previously extracted data. To determine the type of their stenosis, we need more information. By referring to the results shown in Fig. 5, we can obtain a more comprehensive picture of what is happing in the artery.. From Fig. 5-*b* and *c*, it is clear that the maximum stress value of the finger during palpating of cross section B-B is independent of the angle of the rolling process and cross section C-C is dependent on it. This is because the wall thickness of the artery in cross section B-B is identical. So, it can be derived that the stenosis of the artery in cross section C-C is not symmetrical.

**Conclusion**

From Fig. 4-*a*, *b*, and *c*, it can be found that, at a specific time, the stress peak for the stenotic part of the artery is larger than the healthy part. Also, it can be observed that the amplitude of stress variation versus time in the healthy part is larger than the stenotic. Consequently, the separation of the healthy section of the artery from the stenotic is possible by comparing their stiffness and pulse intensity. As shown in Fig. 4-*d* and *e*, the 100% stenotic part of artery and the healthy part without any blood do not have any pulse. The only parameter for differentiating them is the comparison between their stiffness.

According to the results presented in Fig. 5, it can be found that the surgeon feels the same stress value on bottom side of her/his finger in every step of rolling process of healthy (cross sections A-A and E-E) and axisymmetric stenotic (B-B and D-D) parts of the artery. However, this value for stenotic artery is larger than healthy one. But in the part with non-axisymmetric stenosis, the surgeon does not sense the same value of stress as during rolling the artery.

One of the most important features of this research is the benefits of the concept of our results in constructing the surgical simulators for trainees of vascular surgery who can easily obtain adequate information about the existence of the arteries and its stenosis without any kind of penetration into the artery wall. In the current procedures, penetration into the artery wall is necessary during vascular surgery. The other use of these results is its capability of being extended to the field of robotic surgery simulators, where the sense of touch is absolutely necessary for the trainee surgeons to compensate for their inability to palpate the operation sites by long rigid tools.

Further work is currently underway in our laboratory to fabricate a prototype of tactile system to investigate the experimental studies by this prototype for cases in which a tissue includes different arteries and an artery without surrounding tissue. Also, we are working on reducing the size of our tactile system in order to make it portable and to test it clinically.

**ACKNOWLEDGMENT**

The authors gratefully acknowledge Sina Trauma and Surgery Research Center of Iran based in Sina Hospital of Tehran University of Medical Sciences for their supports in this project.